# Ab initio computation of the transition temperature of the charge density wave transition in TiSe$_2$

Dinh Loc Duong[*], Marko Burghard and J. Christian Schön

*Max Planck Institute for Solid State Research, Heisenbergstrasse 1, D-70569 Stuttgart, Germany*

We present a density functional perturbation theory approach to estimate the transition temperature of the charge density wave transition of TiSe$_2$. The softening of the phonon mode at the L-point where in TiSe$_2$ a giant Kohn anomaly occurs, and the energy difference between the normal and distorted phase are analyzed. Both features are studied as function of the electronic temperature, which corresponds to the Fermi-Dirac distribution smearing value in the calculation. The transition temperature is found to be 500 K and 600 K by phonon and energy analysis, respectively, in reasonable agreement with the experimental value of 200 K.

---

[*] To whom correspondence should be addressed; email: l.duong@fkf.mpg.de

# I. INTRODUCTION

Layered materials such as graphite and transition metal dichalcogenides (TMDs) have attracted great interest due to their intriguing electronic and (electro)chemical properties which promise a wide range of applications [1–3]. These materials exhibit weak interlayer interactions along the c direction, as compared to the in-plane bonding, thus facilitating their exfoliation into thin flakes down to monolayer thickness [4,5]. The most interesting feature of these materials is the tunable dependence of their properties on their thickness [6–11]. In addition, the relatively low electronic density of states of an atomically thin TMD flake opens up the possibility to control its properties by gate doping, despite of its (semi)metallic character [5,7]. Moreover, thickness-dependent charge density wave (CDW) transitions in $TiSe_2$ and $TaSe_2$ sheets have recently been experimentally observed [8,10,11].

The CDW state is a typical condensed matter phase of metallic layered TMDs, which has been observed in both the T-phase structure (e.g., 1T-$TiSe_2$, 1T-$VSe_2$, 1T-$TaSe_2$) and the 2H-phase structure (e.g., 2H-$TaS_2$, 2H-$TaSe_2$, 2H-$NbS_2$, 2H-$NbSe_2$) [3]. The associated second order phase transition (called the Peierls transition) is a transformation of a normal stable phase at high temperature to a distorted stable phase at low temperature with a periodic distorted lattice, usually with a larger unit cell [12,13]. The mechanism underlying this transition is the competition between the energy gain due to the gap opening at certain k-points on the Fermi surface and the energy loss due to the lattice distortion in the normal phase [12]. The CDW transition is reversible upon increasing the temperature, since the energy gained by opening the gap is reduced due to the increased number of free thermally excited carriers.

In a metal, the ionic vibrations are strongly screened by the conduction electrons. This screening is decreased at critical q-points, where virtual excitations of the electrons are no longer allowed due to the violation of the crystal momentum conservation law [14]. This leads to a softening of some phonon modes, known as a Kohn anomaly [14]. The amplitude of the Kohn anomaly strongly depends on temperature; in particular, it decreases when the temperature increases due to the smearing of the Fermi surface [15]. This dependence has been explored using density functional perturbation theory (DFPT) simulations in metals, the metallic TMD 2H-NbSe$_2$, carbon nanotubes, graphene and YNi$_2$B$_2$C via tuning the so-called smearing factor [16–22]. This parameter is normally used as a technical tool to achieve convergence for metal systems in DFT calculations. However, when applied in conjunction with the Fermi-Dirac distribution, the smearing factor actually takes on a physical meaning, namely to directly reflect the electronic temperature of the system, and thus describes the occupation probability of the electronic state [23].

In the present context, it is pertinent that the Peierls distortion represents a special case of a Kohn anomaly, where a particular phonon mode is softened down to zero at a transition temperature $T_c$. A transformation of the normal phase to a distorted phase occurs. Although this interplay has been experimentally observed on 2D CDW materials using in-elastic x-ray and neutron scattering, it has not yet been adequately described by *ab initio* approaches, where the occupation probability of the electronic states at a non-zero temperature is not been well described physically [16,24–29]. In this article, we report a detailed study of the CDW transition in TiSe$_2$ by analyzing both, the phonon spectrum of the normal phase and the energy difference relative to the distorted phase, as a function of the electronic temperature via the Fermi-Dirac distribution.

## II. METHODS

DFT calculations were performed using the *Quantum Espresso* (QE) package [30]. The bulk structure of the normal phase of TiSe$_2$ was optimized until the total energy converged within 0.002 eV and the atomic forces on the atoms were smaller than 0.05 eV/Å. The cell structure was fully relaxed until the (internal) pressure was smaller than 0.5 kbar. For the 2x2x2 super-cell structures, a higher convergence of 0.0004 eV in energy and 0.001 eV/Å in force was used. The Fermi-Dirac smearing function was employed as a physical measure of the true electronic temperature of the system [23]. Several sets of irreducible Monkhorst-Pack k-point grid samplings were tested. A projector augmented wave (PAW) potential was implemented with 80 Ry of cut-off energy [31]. Both the local density approximation (LDA) with the Perdew-Zunger functional and the generalized gradient approximation (GGA) with the Perdew-Burke-Ernzerhof functional were used to check the effect of exchange correlation [32,33]. Phonon calculations were also performed by DFPT provided in the QE package [34]. For the phonon dispersion calculation, a 4x4x2 k-point grid was used when calculating the dynamic matrix. The structure of the primitive cell of the normal phase optimized at a smearing value of 789 K was used to calculate the phonon dispersion. To check the independence of the phonon dispersion from the particular smearing value at which the structure was optimized, we repeated the structure optimization for each smearing temperature, followed by a computation of the phonon spectrum at several points of the Brillouin zone, and obtained the same results.

## III. RESULTS

### A. Giant Kohn anomaly and transition from normal phase to CDW phase

The phonon dispersion calculated by GGA of normal phase TiSe$_2$ for different values of the smearing factor with a 24x24x12 k-point mesh is shown in Fig. 1. A narrow soft mode or Kohn anomaly appears at the M and L points of the acoustic phonon branches. The frequencies at these points are changed from real 1.67 (1.77) THz at 789 K to imaginary 1.52i (0.54i) THz at 158 K at the L (M) point. The imaginary frequencies in the phonon dispersion indicate that the structure of the normal phase is unstable at small smearing factors. We note that the pressure in the optimized cell changes only little (below 0.2 kbar) with the smearing value. Such a small pressure change in the unit cell should not induce a notable change in the phonon dispersion curves of TiSe$_2$ [24]. We thus conclude that the imaginary frequencies originate from a purely electronic structure or electron-phonon coupling effect. It should be emphasized that the effect of the smearing factor always needs to be carefully considered when studying the stability of CDW materials based on their phonon dispersion [16,35].

To understand the phase transition deduced from the phonon dispersion analysis in more detail, a 2x2x2 super-cell was studied. The analysis of the phonon spectrum of the 2x2x2 normal phase at the Gamma point reveals three imaginary frequencies at a smearing factor of 158 K, indicating that the structure of the normal phase is unstable under this condition. Fig. 2a illustrates the vibrations of the atoms in these three normal modes. Based upon the corresponding atomic displacements, three distorted initial structures were generated in the super-cell and optimized by energy. All of them return to the normal phase for a smearing value of 789 K, but settle in new distorted structures for a smearing factor of 158 K. This is consistent with the phonon dispersion analysis.

We also observed that a random distortion of the initial structure resulted in another metastable distorted phase, which is clearly distinct from the distorted phases above. The energy difference of these structures is within 5 meV for GGA. The most stable structure of the distorted phase, depicted in the last panel of Fig. 2a, is consistent with conclusions drawn from neutron-diffraction measurements, albeit the atomic displacement is larger (see values in Table 1) [36]. One of the most interesting features is that the displacements of the six Se atoms in the TiSe$_6$ octahedra that belong to the top layer of the 2x2x2 structure (blue arrows) are in the directions opposite to those of the displacements of the Se-atoms in the TiSe$_6$ octahedra in the bottom layer (red arrows). In practice, the distortion can occur along both directions in one TiSe$_2$ layer, thus allowing the formation of CDW domains with different chirality, as has actually been observed by scanning tunneling microscopy [37].

The energies of the normal phase and the distorted phase were calculated for different smearing temperatures with two different k-point meshes (see Fig. 2b). At small smearing temperatures, the distorted phase is stable while at high smearing temperatures the normal phase is stable. Since the energy difference between the two phases is quite small (at T = 158 K, they are 8 meV /supercell and 4 meV/supercell with an 8x8x4 and a 12x12x6 k-point mesh, respectively), we recomputed the energy curve with a higher cut-off energy up to 120 Ry obtaining very similar results. We performed a further check with a Gaussian smearing function, which gives 6 meV of energy gain for a 12x12x6 k-point mesh. We note that the energy gained by the distorted phase (for a smearing of 158 K) in our calculation with the PAW potential is smaller than the 40 meV/supercell (for GGA) given in earlier studies with a mixed potential [28].

The temperature where the two curves in Fig. 2b cross (about 600 K) is in good agreement with the temperature where the softening of the phonon dispersion curves reaches zero in Fig. 1. Our results are also consistent with experimental findings that show a softening of the phonon mode to zero frequency at the L point at the CDW transition temperature [25,27]. We conclude that the CDW phase transition in TiSe$_2$ can be well described within the DFPT framework, employing both the phonon and the energy analysis.

**B. $T_c$ of the CDW phase in TiSe$_2$**

The close relation of the Peierls distortion and the Kohn anomaly is fully mirrored by our results in Figs. 1 and 2. One important feature of the Peierls distortion is the structural change due to a purely electronic effect or electronic temperature. In particular, if the electronic temperature of the system is included in the DFT calculation via a Fermi-Dirac distribution function, a smaller smearing value corresponds to a smaller electronic temperature [23]. Of course, the real temperature affects both ionic and electronic degrees of freedom. Nevertheless, although anharmonic effects due to the motion of the ions at non-zero temperature are not accounted for in our current calculation, it is reasonable to compute a first-order approximation of $T_c$ based on the electronic temperature alone. Thus we have calculated the phonon frequency $\omega_L$ at the L point as function of the electronic temperature converted from the smearing value (see Fig. 3). Assuming that mean-field theory is a good first approximation for the description of the second-order phase transition, we fit the data according to the functional form $\omega_L(T) = \omega_0*(T-T_c)^\delta$. As in real systems imaginary frequencies are not physical for stable configurations, we only include the real $\omega_L$ values in the fit. The results are listed in Table 2. For bulk TiSe$_2$ calculations with the GGA functional, the best fit is obtained for $T_c$ = 503 K with a 24x24x12 k-point mesh. The estimated value of $T_c$ is in good agreement with the value estimated from the analysis of the energy difference between normal and distorted phases and reasonably close to the experimental value of 200 K. The error in $T_c$ is most likely due to the limitations in the number of k-points, the anharmonic effects of ionic motion and the exchange correlation functional (see below). We note that usually the phonon becomes stiffer for lower temperatures, and thus the computed $T_c$ values should represent an upper bound. The power-law exponent is quite close to the value of 0.5 found in the mean-field approximation in a free electron model [15].

## IV. DISCUSSION

### A. Effects of k-point grids on $T_c$

In the preceding section, we have shown that an analysis of the phonon spectrum as function of temperature, together with an investigation of the energies of normal and distorted phases, allows identifying CDW-transitions in a material, and provide reasonable estimates of the corresponding transition temperatures. We note from Fig. 3 that an increase in the size of the k-point mesh leads to lower values of $T_c$, but different types of k-point grids give rise to a variation in the $T_c$-values. These estimates are expected to reach a smaller value $T_c^\infty$ in the limit of an infinite mesh. Of course, $T_c^\infty$ will most likely still be different from the experimentally observed value. Therefore, the value of $T_c \approx 500$ K is only an upper bound of the true $T_c$. In principle, an estimate of the infinite-mesh limit could be obtained by additional calculations with much larger k-point grids; however, such denser meshes would require very large computational resources. For the same reason, we did not compute the energy difference between the normal and distorted phase for (smearing) temperatures near 0 K, since accurate calculations with the Fermi-Dirac distribution would require very large k-point meshes.

### B. Effects of smearing function on $T_c$

In this context, we stress the importance of employing the Femi-Dirac function to physically describe the probability of electronic state occupations. Other functions, which are commonly used to obtain fast convergence in DFT simulations, can easily return an overestimation of $T_c$. In fact, we have repeated the calculations with Gaussian, Marzari-Vanderbilt and Methfessel-Paxton functions, which resulted in a much larger $T_c$ than for the Fermi-Dirac function, as shown in Fig. 4a-c. The imaginary frequencies appear between 1580 K and 789 K of smearing values for a Gaussian function. The transition point is even larger than 1580 K in the case of the Marzari-Vanderbilt and Methfessel-Paxton functions. The same phenomenon has been reported in NbSe$_2$ where a Gaussian smearing function was used [16]. The convergence of $T_c$ with k-point grids using the Gaussian function is shown in Fig. 4d. The same behavior of the variation of $T_c$ with the type of k-point grids is observed as in the case for the Fermi-Dirac smearing function. The 12x12x6 and 24x24x12 meshes, which belong to the same type of the k-point grid, give a lower value compared to the 16x16x8 k-point grid.

### C. Effect of the exchange correlation functional

The phonon dispersion curves of TiSe$_2$ calculated by LDA with different sets of k-point grids are shown in Fig. 5. They are strongly sensitive to the choice of k-point grid, much more than the curves computed using GGA. Although a giant Kohn anomaly appears at the M and L points in all curves, a transition from a real to imaginary value at the L point takes place only for the smallest 16x16x8 grid. No transition is observed for denser k-point grids. This strong k-point dependence was also observed when using other smearing functions as shown in Fig. 6. We also optimized the structure of a distorted 2x2x2 double cell with 158 K of smearing. While stable distorted structures are obtained with an 8x8x4 mesh (a half of the 16x16x8 mesh used in the primitive cell), they relaxed to the normal phase already with a 12x12x6 mesh (a half of the 24x24x12 mesh used in the primitive cell), consistent with the phonon analysis when using the LDA functional.

Basically, the driving force behind the CDW transition is the opening of a band gap in the electronic density of states, with an associated reduction in the electronic energy. Thus in order to explain the disagreement of the GGA and LDA calculations, and in particular the failure of the LDA to show a CDW transition for large k-point meshes, the band structures of the normal and distorted phases obtained with the 8x8x4 k-point mesh in LDA and GGA along the Γ-M and A-L paths (shown in Figs. 7 and 8) have been analyzed. There are several interesting changes in the band structure of the distorted phase compared to the one of the normal phase (marked by olive color rectangles as regions 1 and 2 in Fig.7). The slopes of the band dispersion are smaller in the distorted phase, and there are new occupied states in region 1 and new empty states on the top of region 2. This indicates a net-redistribution of electrons to lower energy levels with an associated gain in energy. These changes are much bigger in GGA than in LDA. A similar phenomenon occurs along the A-L path (Fig. 8). We thus suggest that the failure of the LDA-calculations to consistently reproduce the Peierls-transition in $TiSe_2$ is due to the fact that the gain in energy usually caused by the net-redistribution of electrons into new states with lower energy levels is not well realized in LDA.

Our calculations show that the mechanism of the CDW transition in $TiSe_2$ is essentially due to an electronically driven mechanism or electron-phonon coupling, in accordance with recent experiments [38]. However, the demonstration of the existence of the transition based on phonon curves is strongly dependent on the choice of exchange-correlation functionals. While GGA clearly shows the transition, LDA fails to predict this transition. In this context, we note that the question of the possible role of excitonic effects that have been suggested to play a role in the CDW-transition of $TiSe_2$, cannot be addressed by density functional theory in the framework of either LDA or GGA, pointing to the necessity of further investigations to clarify this issue.

**D. Kohn anomaly at M point**

In addition to the Kohn anomaly at the L point, we observed a smaller Kohn anomaly at the M point. The frequency of the soft mode at the M point is always real when using the LDA functional. In contrast, the frequency becomes imaginary for all smearing functions when using the GGA functional, except for the Marzari-Vanderbilt function. The 2x2x2 distorted phase observed by experiments corresponds to the instability of the soft mode only at the L point, not at the M point. The disagreement between the experimental result and the theoretical prediction may originate from an underestimation of the Van-der-Waals interaction between layers along the c-axis using the GGA functional. This underestimation lowers the coupling between layers, giving rise to the instability at the M point.

**E. General applicability of the approach**

As mentioned above, we have employed two ways to estimate the value of $T_c$. Although the value of $T_c$ can be directly estimated from the energy curves shown in Fig. 2b, the approach based on the phonon dispersion curves has certain advantages. Most importantly, the normal phase with its small unit cell can be used for computing the phonon dispersion, thus avoiding the much more expensive calculations usually necessary for commensurate and incommensurate CDW phases. For example, using a super-cell method is impossible in the case of an incommensurate distorted phase, due to the requirement of an essentially infinite super-cell. Furthermore, it is not always trivial to identify the correct distorted structure in the super-cell unless information about the soft modes in the system are available - as we have shown above, a random displacement away from the normal phase does not necessarily lead to the proper minimum energy structure.

To demonstrate the more general applicability of our approach to study the CDW transition, we have investigated whether the transition in $TiSe_2$ is also present at high pressures. To this end, the phonon dispersion at a pressure of 5 GPa was calculated with the GGA functional (see Fig. 9). The frequency of the soft mode at the L point remains real for all smearing values down to 1.5 K. We notice that the frequency of this soft mode monotonically decreases with decreasing smearing value as long as the smearing is larger than 158 K. Below this value, the frequency of the soft mode actually increases with decreasing smearing value. This indicates that the normal phase is stable at any electronic temperature at the pressure of 5 GPa, consistent with experimental results [39].

## IV. CONCLUSIONS

In summary, we have investigated the effects of the electronic temperature on the Kohn anomaly and energy of $TiSe_2$ in a 2x2x2 super-cell, via *ab initio* simulations based on DFPT. The CDW phase transition manifests itself in phonon softening of the dispersion curves of the normal high-temperature phase, along with the crossing of the energies of the normal and distorted phase as function of electronic temperature. As a major finding, the transition temperature to the CDW phase can be estimated with reasonable accuracy based on the electronic temperature using the Fermi-Dirac distribution for the occupation of the electronic energy levels, which gives a value on the same order as observed by experiments. This establishes a generic approach of fast *ab initio* level prediction of the effects of thickness, strain and doping on the CDW phase transition temperature for a wide range of quasi-2D layered transition metal compounds.

**ACKNOWLEDGMENTS**

We gratefully acknowledge R. Gutzler, P. Horsch, V. Olevano, J. Skelton, and A. Walsh for valuable discussions. The calculations were performed on a high performance cluster of the Karlsruhe Institute of Technology (KIT) within the ELCPHON project.

**Figure captions**

FIG. 1. Phonon dispersion of normal phase $TiSe_2$ computed with GGA functional, for several smearing factors using the Fermi-Dirac function as smearing function with a 24x24x12 k-point mesh. Giant Kohn anomalies are visible at the M and L points. They change from real to imaginary values when the smearing values decrease 789 K to 158 K.

FIG. 2. (a) Three distorted structures predicted by normal mode analysis of the imaginary frequencies at the L point of a 2x2x2 normal phase at a smearing value of 158 K. The energy difference of these structures with respect to the structure with lowest energy (structure on the right) is listed on the top. The displacement (exaggerated for visibility) of the atoms of the $TiSe_6$-octahedra in the top (bottom) layer is indicated by blue arrows (red arrows). The rightmost configuration is the most stable structure among the distorted configurations, in agreement with experimental observations [36]. (b) The energy of the 2x2x2 normal phase (black) and distorted phase (red) with different smearing values computed using GGA with 8x8x4 (left plot) and 12x12x6 (right plot) k-point mesh. The energy of the normal phase at a smearing of 0.001 is taken as the reference and set to zero.

FIG. 3. Frequency $\omega_L$ at the L point plotted as a function of simulated electronic temperature with GGA functional for four different k-point grids. The solid line is the mean-field fit according to $\omega_L(T) = \omega_0*(T-T_c)^\delta$. Here, $T_c$ is the transition temperature and $\delta$ is the exponent, varying from 0.48 to 0.53 for different meshes, which is quite close to the value of 0.5 for mean-field models.

FIG. 4. Phonon dispersion curves of TiSe$_2$ for alternative smearing functions: Gaussian (a), Marzari-Vanderbilt (b), Methfessel-Paxton (c). In all cases, a 24x24x12 k-point grid was used with the GGA functional. The estimated T$_c$ from the Gaussian function lies between 789 to 1580 K and is higher than 1580 K when using the Marzari-Vanderbilt and Methfessel-Paxton smearing functions. All three are much higher than the value estimated for the Fermi-Dirac function. (d) The estimated T$_c$ using the Gaussian smearing function for three different k-point grids. The lowest T$_c$ is around 1420 K.

FIG. 5. The phonon dispersion of normal phase TiSe$_2$ computed by the LDA functional with several smearing factors using the Fermi-Dirac smearing function for the 16x16x8, 24x24x12 and 32x32x16 k-point meshes. The frequency of the soft mode changes to an imaginary value only in the case of the 16x16x8 k-point mesh. The frequency of soft modes remains real even with very small smearing values (down to 15 K) for the denser k-point grids.

FIG. 6. The phonon dispersion of normal phase TiSe$_2$ computed with the LDA functional for several smearing factors using three alternative smearing functions (Gaussian, Marzari-Vanderbilt, Methfessel-Paxton) for the 16x16x8, and 24x24x12 k-point meshes. The frequency of the soft mode changes to imaginary value only for the coarse 16x16x8 k-point mesh. However, no imaginary frequency is observed at a denser 24x24x12 k-point mesh for any smearing temperature.

FIG. 7. The band structure from the Gamma to M point of the normal and distorted

phases (obtained using a 8x8x4 k-point mesh for a 2x2x2 cell) calculated using GGA (left) and LDA (right). The regions where the differences between the normal and the distorted phase are most pronounced are marked by olive color rectangles.

FIG. 8. The band structure from the A to the L point of the normal and distorted phases (obtained using a 8x8x4 k-point mesh for a 2x2x2 cell) calculated using GGA (left) and LDA (right). The regions where the differences between the normal and the distorted phase are most pronounced are marked by olive color rectangles.

FIG. 9. The phonon dispersion of $TiSe_2$ at 5 GPa computed with the GGA functional for a k-point mesh of 16x16x8. No imaginary frequency is observed. The CDW phase of $TiSe_2$ is suppressed at this pressure.

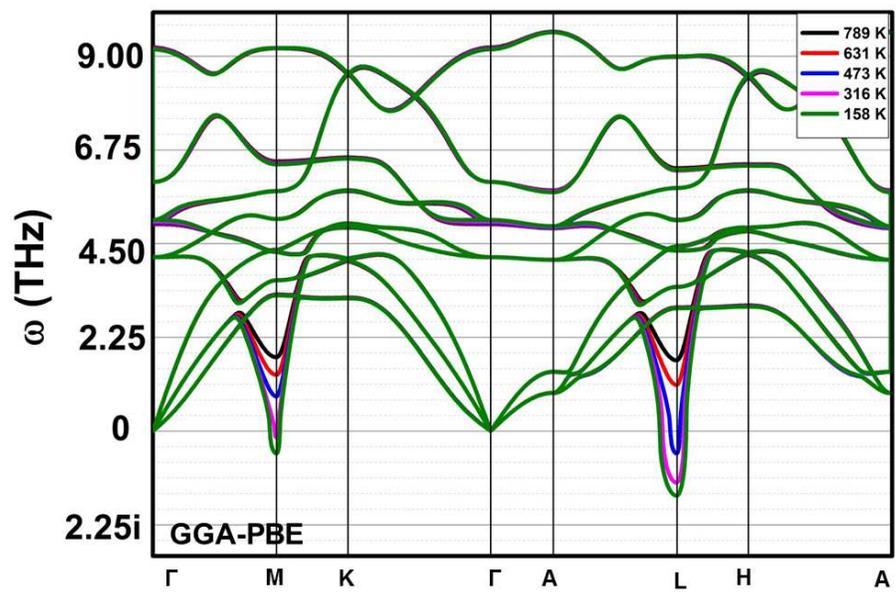

FIG. 1

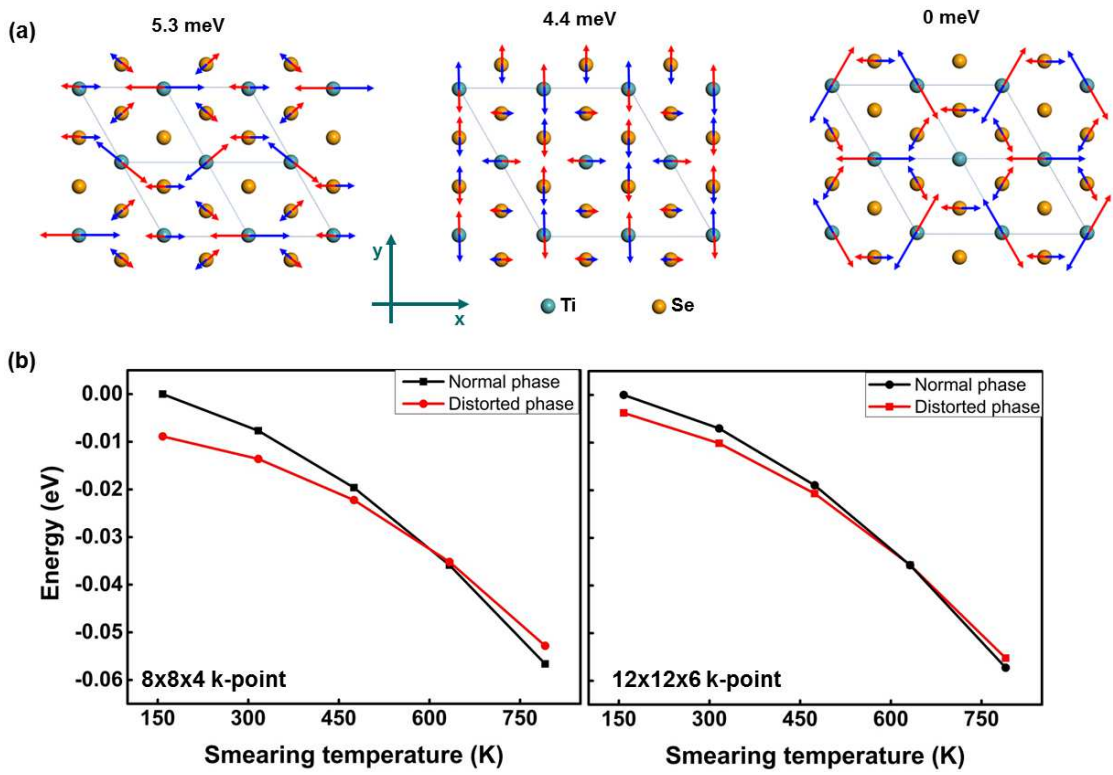

FIG. 2

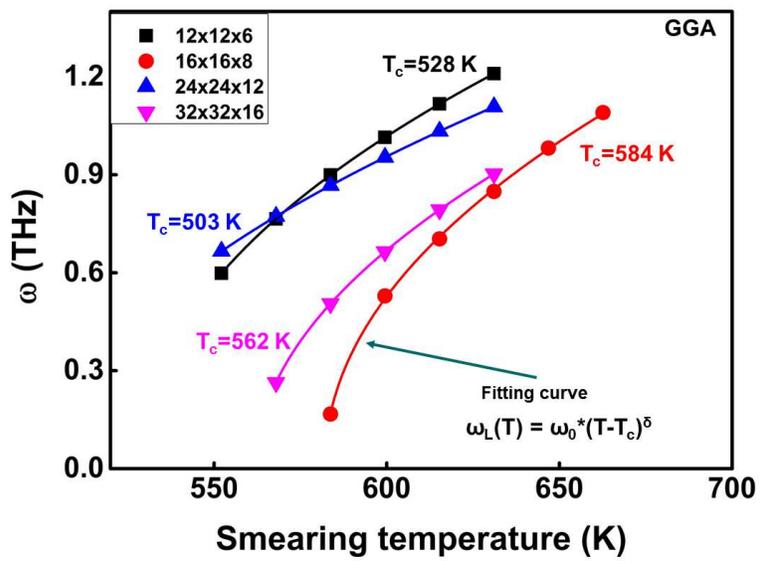

FIG. 3

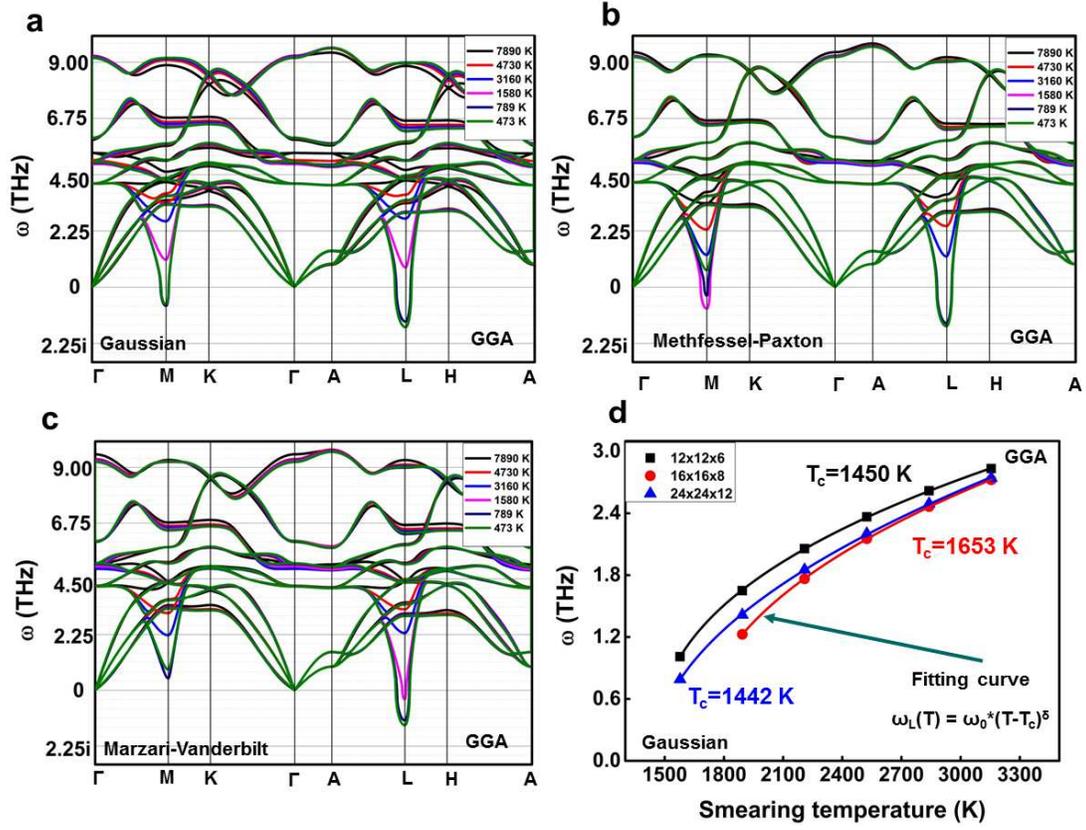

FIG. 4

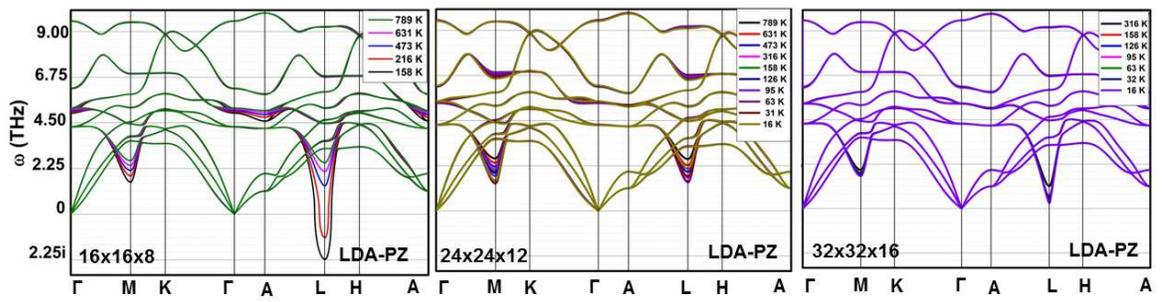

FIG. 5

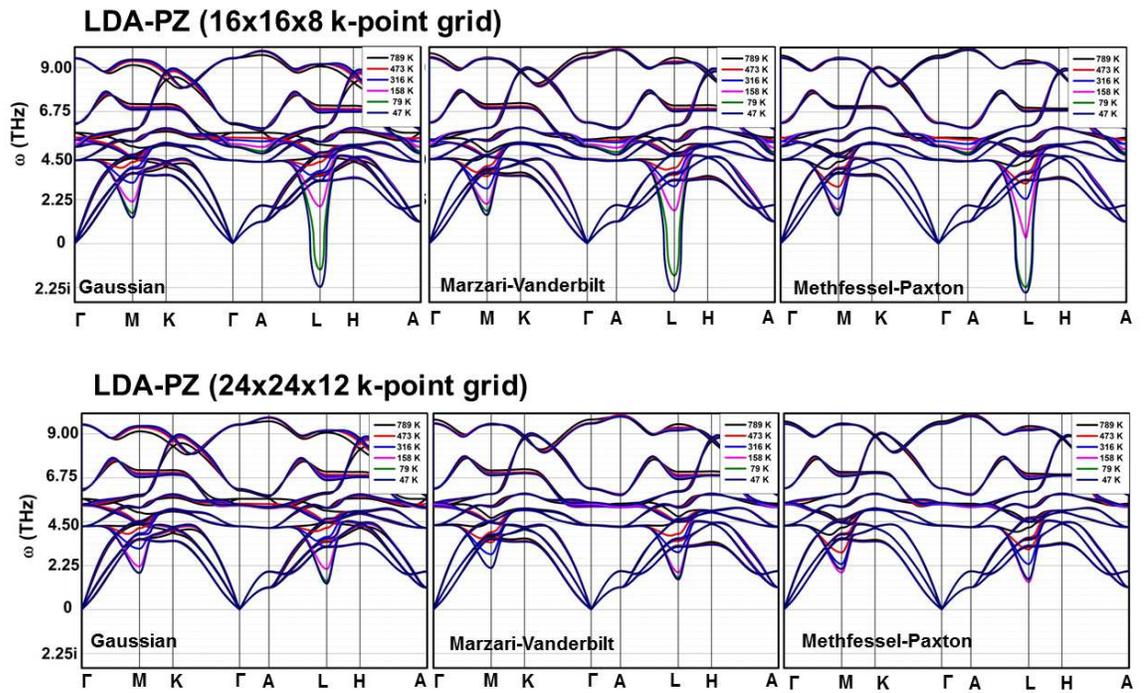

FIG. 6

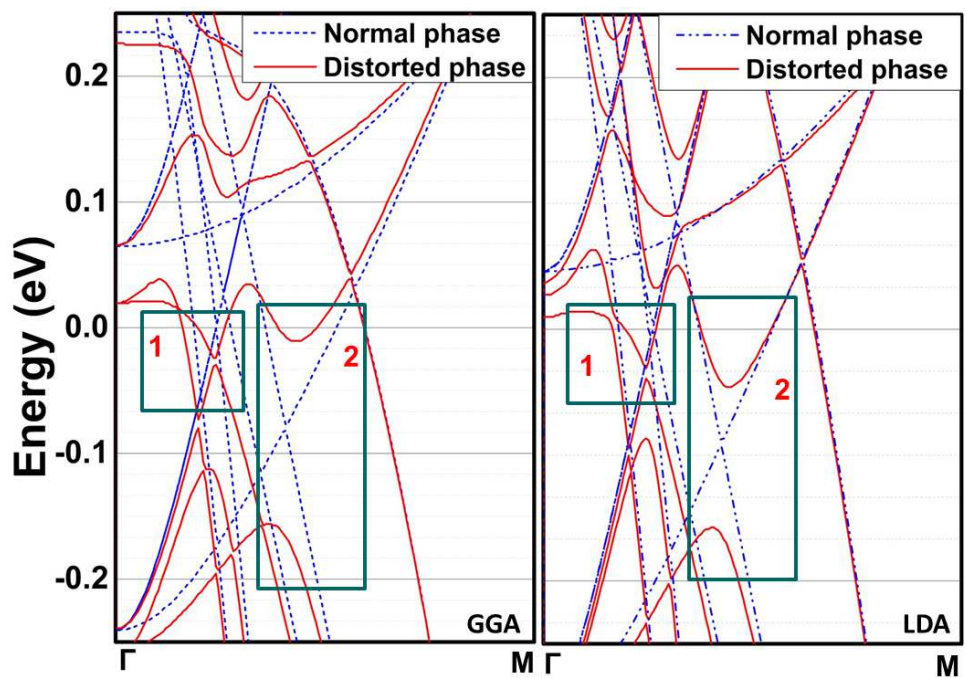

FIG. 7

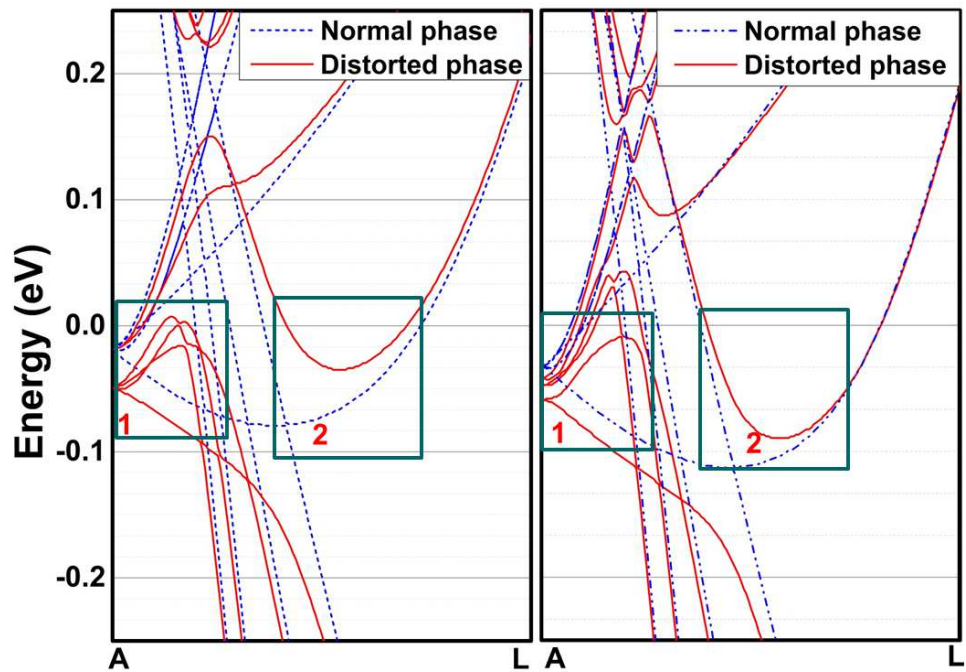

FIG. 8

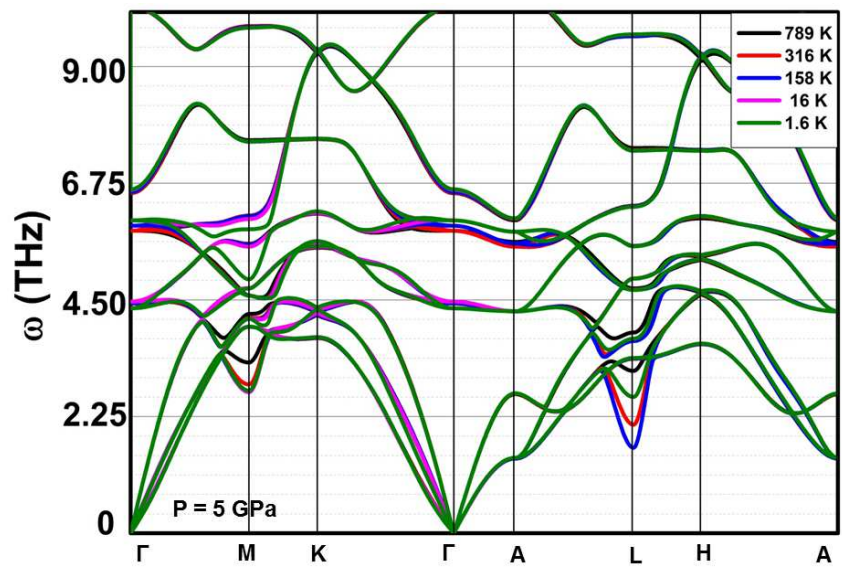

FIG. 9

Table 1: Displacements $\delta_{Ti}$ and $\delta_{Se}$ of Ti and Se atoms with respect to their positions in the normal high temperature phase observed in the most stable distorted low temperature structure, given in units of Ångstrom and the lattice constant a.

|      | $\delta_{Ti}$ (Å) | $\delta_{Ti}/a$ | $\delta_{Se}$ (Å) | $\delta_{Se}/a$ | $\delta_{Ti}/\delta_{Se}$ |
|------|-------------------|-----------------|-------------------|-----------------|---------------------------|
| GGA  | 0.091             | 0.026           | 0.035             | 0.010           | 2.6                       |
| Exp* | 0.042             | 0.012           | 0.014             | 0.004           | 3.0                       |

*Data extracted from ref. [36]

Table 2: Fitting parameters of Fig.3

|          | $T_c$ (K) | $\omega_o$ (THz) | $\delta$ |
|----------|-----------|------------------|----------|
| 12x12x6  | 528       | 0.13             | 0.477    |
| 16x16x8  | 584       | 0.13             | 0.481    |
| 24x24x12 | 503       | 0.08             | 0.529    |
| 32x32x16 | 562       | 0.11             | 0.504    |